\documentclass[aps,prl,twocolumn,superscriptaddress,preprintnumbers]{revtex4}
\usepackage{graphicx}
\usepackage{amsmath,amssymb,bm}
\usepackage[mathscr]{eucal}
\usepackage[colorlinks,linkcolor=blue,citecolor=blue]{hyperref}

\newif\ifarxiv
\arxivfalse

\begin		{document}
\def\gsim{\mbox{~{\protect\raisebox{0.4ex}{$>$}}\hspace{-1.1em}
	{\protect\raisebox{-0.6ex}{$\sim$}}~}}
\def\lsim{\mbox{~{\protect\raisebox{0.4ex}{$<$}}\hspace{-1.1em}
	{\protect\raisebox{-0.6ex}{$\sim$}}~}}
\def\st    {\begin{equation}}
\def\stp    {\end{equation}}

\def\Nfour	{\mathcal N\,{=}\,4}

\def\T		{\mathcal T}
\def\q		{\bm q}

\def\half	{{\textstyle \frac 12}}

\def\Arxiv      #1 [#2]{\href{http://arxiv.org/abs/#1}{{\tt arXiv:#1 [#2]}}\,}

\def\half{{\textstyle\frac{1}{2}}}
\def\Fig#1{Fig.~\ref{#1}}
\def\Eq#1{eq.~(\ref{#1})}

\title
    {
    Dilaton emission and absorption from far-from-equilibrium non-abelian plasma 
    }

\author{Paul~M.~Chesler}

\affiliation
    {%
Department of Physics, 
MIT, 
Cambridge, MA 02139, USA 
    }%

\author{Derek~Teaney}

\affiliation
    {%
    Department of Physics \& Astronomy,
    SUNY at Stony Brook,
    Stony Brook, NY 11794, USA
    }%

\date{\today}

\begin{abstract}
Using gauge/gravity duality, we study emission and absorption rates of scalar quanta
from far-from-equilibrium $\Nfour$ supersymmetric Yang-Mills plasma.  
By comparing the emission and absorption rates 
to expectations from the Fluctuation-Dissipation Theorem, we study how 
the spectrum thermalizes and how the thermalization time depends
on the four momentum of the emitted quanta. 
\end{abstract}

\preprint{MIT-CTP-4413}


\maketitle

{\it{Introduction}}.--Gauge/gravity duality \cite{Maldacena:1997re}, or
holography, is a powerful tool for studying real-time dynamics in strongly
coupled quantum field theories.  Through holography the creation and
thermalization of non-abelian plasma maps onto the process of gravitational
collapse and black hole thermalization, which can be studied numerically.
The equilibration of holographic plasma can yield valuable insight
into the thermalization of Quark-Gluon Plasma (QGP) produced
in heavy ion collisions.
In addition, precise questions about thermalization in
the quantum field theory can clarify the statistical properties of
non-equilibrium black holes.
The simplest theory to study with a gravitational dual is
 $\Nfour$ supersymmetric Yang-Mills theory (SYM) at large $N_{\rm c}$
and 't Hooft coupling $\lambda$.

As analog problem of photon production in  the QGP, we study the
emission of weakly coupled ($4d$) dilatons from strongly coupled non-equilibrium SYM
plasma.
Dilaton emission and absorption rates are proportional to the Fourier transform of the
Wightman functions,
 $g_-(x |x') = \langle \hat O(x') \hat O(x) \rangle$ and 
$g_+(x| x') = \langle \hat O(x) \hat O(x') \rangle$  respectively, 
where $O(x)$ is the SYM Lagrange density.
In thermal equilibrium at temperature $T$ these rates are related by the Fluctuation Dissipation Theorem (FDT) 
\st
\label{eqFDT}
     g_-(q) = e^{-\omega/ T}  g_+(q)  \, , 
\stp
where $q\equiv(\omega, \q)$ is the four-momentum conjugate to $x-x'$.
The FDT is a direct consequence of the fact that the density 
matrix of the system takes its equilibrium form  $e^{-H/T}$. 
Therefore, the degree to
which the FDT is satisfied in non-equilibrium plasmas provides an  unequivocal
measure of the equilibration a particular Fourier mode $q$.

Through holography, the emission and absorption of weakly coupled quanta from non-equilbrium SYM plasma is dual to the emission and absorption of Hawking radiation from a non-equilibrium $5d$ black brane. 
More precisely, the $4d$ SYM Wightman functions $g_\pm$
are encoded in $5d$ Wightman functions $G_\pm$
computed in the black brane geometry.  
The Wightman functions satisfy classical equations of motion 
which can be solved numerically in non-equilibrium geometries after
the appropriate quantum mechanical initial conditions are specified \cite{CaronHuot:2011dr}.  
Only the  absorption rate of a classical $4d$ dilaton field, $g_+ - g_-$, is determined by the classical response of the black brane \cite{deBoer:2008gu,Son:2009vu}.

For definiteness,  we will study the thermalization of
homogeneous but anisotropic SYM plasma.  
To create this  non-equilibrium state,  
we consider the response of an initially equilibrium SYM plasma at temperature $T_{\rm i}$ to a 
time-dependent gravitational field \cite{Chesler:2008hg}.  The homogeneous $4d$ metric of the external field  is
\begin{equation}
\label{boundarygeometry}
ds^2 = -dt^2 + e^{b(t)} \, d \bm x_{\perp}^2 + e^{-2 b(t)} \, dx_{||}^2 \,,
\end{equation}
where 
$\bm x_{\perp} \equiv \{x_1,x_2\}$,  
and the time dependence of the field is 
Gaussian $b(t) = c/(\sqrt{2 \pi \T^2}) e^{-t^2/2\T^2 }$. 
The time-dependent
geometry does work on the SYM plasma and creates a non-equilibrium state
with an anisotropic stress $\langle T^{\mu \nu} \rangle = {\rm diag} [ \mathcal E, \mathcal P_{\perp},\mathcal P_{\perp},\mathcal P_{||}]$, which
subsequently equlibrates at late times.
The amplitude and width of $b(t)$, $c = 2.11, {\mathcal T}  =
1/\pi T_{\rm f} $, are adjusted using the simulations described below so that the energy density  changes by a
factor of fifty,  and the final temperature is $T_{\rm f} = 50^{1/4} T_{\rm
i}$. 

We will quantify and analyze the equilibration of system by monitoring the emission and absorption rates in a given frequency band as a function of time. 
Clearly, the frequency and time resolutions of this analysis are limited by the 
uncertainty principle, $\Delta \omega\Delta t \geq \half$.
To achieve the best possible  resolution in both frequency and time
we will compute Wigner transforms of $g_{\pm}(x,x')$
averaged over phase-space with a minimal uncertainty wave packet. 
The Wigner transforms
are 
\begin{equation}
\label{wigtrans}
g_{\pm}(\bar t,\omega,\q) \equiv \int d (\Delta x) \, g_{\pm}(x|x') e^{-i q \cdot \Delta x},
\end{equation}
with $\Delta x = x - x'$ and $\bar t = \frac{1}{2}(t + t')$,  and 
the minimally averaged transforms, known as Gabor transforms, are
\begin{equation}
   \label{smeared_correlator}
\bar g_\pm(\bar t, \omega,\bm q)  \equiv \int \frac{d\bar t' d\omega'}{2\pi}  
e^{-\Delta \omega^2 \sigma^2} e^{-\Delta \bar t^2/\sigma^2 } g_\pm(\bar t',\omega',\bm q) \, , 
 \end{equation}
 where $\Delta\omega = \omega' - \omega$,  $\Delta \bar t = \bar t' - {\bar t}$,
 and  $\sigma$ is a temporal resolution parameter which we choose to be $\sigma=1/\pi T_{\rm f}$.
For the homogeneous system under study, 
the energy density is constant once the external metric  
is static,  and  
an effective temperature can be defined via the energy density
\begin{equation}
   \frac{1}{\beta_{\rm eff}(t) }\equiv T_{\rm eff}(t) \equiv  \left (8 |\mathcal E(t)|/(3 \pi^2 N_{\rm c}^2) \right)^{1/4} \, .
\end{equation}
If $g_\pm(q)$ satisfies the FDT at late times, then the Gabor transforms satisfy
\begin{equation}
   \label{smearedFDT}
   \bar g_-(\bar t,\omega, \bm q) = e^{-\omega \beta_{\rm eff}(\bar t) + \beta^2_{\rm eff}(\bar t) /4\sigma^2}  \bar g_+(\bar t, \omega - \beta_{\rm eff}(\bar t)/2\sigma^2, \bm q) \, .
\end{equation}
The degree to which (\ref{smearedFDT})  is satisfied provides an excellent definition of equilibration.

\textit{Gravitational description.}---The $5d$ gravitational dual to an anisotropic SYM plasma created
by a time dependent $4d$ geometry was studied in \cite{Chesler:2008hg}.  We follow the analysis 
of \cite{Chesler:2008hg} and outline some of the salient features.

Following \cite{Chesler:2008hg}, using translation
invariance we may write the $5d$ metric as
\begin{align}
\label{metric} 
ds^2 = &-A \, dt^2 + \Sigma^2 \big [ e^{B} d \bm x_{\perp}^2 +
e^{-2 B} dx_{||}^2 \big ] + 2 dr \,dt\,, 
\end{align} 
where $A$, $B$, and
$\Sigma$ are functions of $r$ and $t$, only.
The coordinates $t$ and $r$ are generalized Eddington-Finkelstein coordinates.
Infalling radial null geodesics have constant values of $t$ (as well as $\bm
x_\perp$ and $x_{||}$). 
The $4d$ boundary of the $5d$ geometry is located at $r = \infty$, 
and is proportional to the geometry of the SYM plasma in \Eq{boundarygeometry}.

The  evolution of the $5d$ metric (\ref{metric}) is determined by solving 
numerically Einstein's equations with negative cosmological constant
\cite{Chesler:2010bi}.
Demanding that the boundary metric is that of (\ref{boundarygeometry}), equates
to imposing the boundary conditions,
$\lim_{r \to \infty} B(t,r) = b(t), \ \lim_{r \to \infty} \Sigma(t,r)/r = 1 $
\cite{Chesler:2008hg}.
The near-boundary behavior of the metric determines the expectation value
of the SYM stress $\langle T^{\mu \nu} \rangle$ \cite{deHaro:2000xn}.

In the infinite past, when $b(t) = 0$, we take the initial geometry to be an
equilibrium black brane of temperature $T_{\rm i}$.
When the boundary source turns on,  $b'(t) \neq 0$,
the changing boundary geometry creates 
gravitational radiation that propagates into the bulk creating a non-equilibrium black brane. 
When the boundary geometry is static, $b'(t)  = 0 $, the $5d$ geometry relaxes
to an equilibrium black brane with temperature $T_{\rm f}$.  
After the solving the Einstein equations, 
the event horizon $r_{\rm h}(t)$ can be found by tracing null radial geodesics 
 backwards in time  from the far future.


As the black brane evolves and equilibrates, it continuously emits and absorbs
Hawking radiation.
The radiation 
propagates up from the horizon to the boundary, where it is reflected back
towards the horizon and is absorbed. 
The process continues until the $5d$ dilaton modes are in equilibrium with the
black brane and are thermally occupied with distribution function $n(\omega) = 1/(e^{\omega/T_{\rm f} }-
1)$.
This competitive evolution between emission and absorption  is
the gravitational dual of thermalization in SYM plasmas.


The emission of $5d$ dilatons is encoded in the $5d$ Wightman function
$G_{-}(x_1,r_1|x_2,r_2)   = \langle \hat \varphi(x_2,r_2)\hat\varphi(x_1,r_1)
\rangle$ where $\varphi$ is the dilaton field.  Similarly, the absorption of
$5d$ dilatons is encoded in $G_{+}(x_1,r_1|x_2,r_2)   = \langle
\hat\varphi(x_1,r_1)\hat\varphi(x_2,r_2) \rangle$.
Both $G_{\pm}$ satisfy the classical equations of motion 
\begin{equation}
\label{homoeqm}
-D_{(1)}^2 G_{\pm} (x_1,r_1|x_2,r_2) = - D_{(2)}^2 G_{\pm}(x_1,r_1|x_2,r_2)  = 0,
\end{equation}
where $D^2_{(n)}$ is the covariant wave operator under the background geometry 
with respect
to coordinates $\{x_n,r_n\}$.

Previously,  we (together with S. Caron-Huot) formulated a tractable 
numerical recipe for solving eqs.~(\ref{homoeqm}) \cite{CaronHuot:2011dr}.  
Our prescription is based on the following logic. First, we note
that $G_{\pm}$ satisfies the homogeneous
equations of motion (\ref{homoeqm}) and is not a Green function. 
Thus, after specifying the initial conditions in infinite past,
forward evolution determines $G_{\pm}$ in
the future. In the equilibrium black brane geometry of the distant past,
most of the initial data specified for $G_{\pm}$ falls through the black brane horizon in a
finite time and thus becomes irrelevant for future evolution.  Examining
the geodesics of the equilibrium black hole geometry, one concludes that the only
initial data of relevance to the evolution of $G_{\pm}$ in the future is determined by the coincident point singularities of
the two point functions arbitrarily close to the event horizon, {\it i.e.}  ultraviolet vacuum
fluctuations. 
These UV vacuum fluctuations propagate along horizon
skimming geodesics and eventually emerge (redshifted) from the stretched horizon at $r_* = r_h(t) + \epsilon$.  
The flux of data passing through the stretched horizon
can be traded for effective sources defined on the stretched horizon and the solution to (\ref{homoeqm}) can be written
in terms of the convolution integral 
\begin{align}
\label{bulk2bnd}
\nonumber
G_{\pm}(x_1,r_1|x_2,r_2)  = \!\!\!\! \int\limits_{r'_n=r_{*}}  \!\!\!\!  \Big \{  & G_{\rm R}(x_1,r_1|x'_1,r'_1) 
 G_{\rm R}(x_2,r_2|x_2',r_2')
 \\ 
 \times &\, {\mathcal G}^{\rm h}_{\pm}(x_1'|x'_2) \Big \},
\end{align}
where the \textit{horizon correlators} ${\mathcal G}^{\rm h}_{\pm}$ are the effective sources and 
the retarded Green's function $G_{\rm R}$ satisfies  
\begin{equation}
\label{greens}
-D_{(1)}^2 G_{\rm R}(x_1,r_1|x_2,r_2) = \frac{\delta^4(x_1 - x_2) \delta(r_1 - r_2)}{\sqrt{-g(x_1,r_1)}},
\end{equation}
with $g$ the determinant of the metric.

The horizon correlators can be computed analytically \cite{CaronHuot:2011dr}.  Specifically, the propagation of the UV vacuum
fluctuations from arbitrarily close to the event horizon to the stretched horizon can be treated using geometric optics.
Furthermore, the UV vacuum fluctuations generated by coincident point singularities 
are universal and independent of initial conditions for $G_\pm$.
Following the procedure of \cite{CaronHuot:2011dr}, in the $\epsilon \to 0$ limit we find
\begin{multline}
\label{horizoncorrelators}
\mathcal G^{\rm h}_{\pm}(x|x')  =  \frac{K \delta^{3}(\bm x - \bm x')}{\left [g(t,r_h) g(t',r_h) \right ]^{1/4} }\bigg \{\pm   i \delta'(t - t')
\\
  -{\textstyle \frac{1}{4 \pi}} \kappa(t) \kappa(t') {\rm csch}^2 {\textstyle \frac{\tau(t')  - \tau(t)}{2}}
 \bigg \},
\end{multline}
where $\kappa(t) \equiv \frac{1}{2} \partial_r A(t,r)|_{r = r_h}$, $\tau(t) \equiv \int^t dt' \kappa(t')$, and $K$ is the normalization of the dilaton action,
$S=-\frac{K}{2}\int \sqrt{-g} (\partial \phi)^2$.  
Once the bulk correlators are determined, the corresponding SYM correlators are given by \cite{CaronHuot:2011dr}
\begin{equation}
g_\pm(x|x') =  \lim_{r,r' \to \infty} {16}\, r^4 r'^4  G_\pm(x,r|x',r').
\end{equation}

The numerical calculation of 
$G_{\pm}$ and hence $g_\pm$ can be greatly simplified by noting that the horizon correlators can be written
\begin{multline}
\mathcal G^{\rm h}_{\pm}(x|x') =  K \frac{ \kappa(t) \kappa(t') \delta^3(\bm x - \bm x')}{[g(t,r_h)g(t',r_h)]^{1/4}} 
\\
\times \int \frac{2\nu d \nu}{2 \pi}  n_{\pm}(2\pi \nu)  e^{-i \nu (\tau(t) - \tau(t'))},
\end{multline}
where $n_{-}(x) = 1/(\exp(x) - 1)$ and $n_{+} = 1 + n_-$ are the Bose 
occupation and stimulation factors respectively.
From eqs.~(\ref{bulk2bnd}) and (\ref{greens}) it follows that the bulk correlators are
\begin{align}
G_{\pm}(x,r|x',r') =  K \! \int \! \frac{2\nu d \nu}{2 \pi} n_\pm(2\pi \nu)  F_\nu(x,r) F^*_\nu(x',r') \, ,
\end{align}
where $F_\nu(x,r)$ satisfies
\begin{equation}
\label{feq}
-D^2 F_\nu(x,r) = \frac{\kappa(t) e^{-i \nu \tau(t)} }{[-g(t,r_h)]^{1/4}} \delta(r- r_*) \delta^{3}(\bm x - \bm x') \, .
\end{equation}

Using translation invariance in the spatial directions, we introduce 
a spatial Fourier transform and solve \Eq{feq} for the mode functions $F_\nu(t,\bm q,r)$.
The required boundary conditions needed to solve \Eq{feq} are that $F_\nu(t,\bm q,r) \to 0$ 
at the boundary and that $F_\nu(t,\bm q,r)$ is regular at the horizon.  The initial conditions 
for $F_\nu(t,\bm q,r)$ 
in the infinite past can be computed from an equilibrium analysis of \Eq{feq} for a static black brane at temperature $T_{\rm i}$.
Finally, we note that 
using field redefinitions it is possible to take the $r_* \to r_{\rm h}$ limit in our numerical analysis.
Further numerical details will be presented elsewhere.

\begin{figure}
\includegraphics[width=0.48\textwidth,height=0.85\textheight]{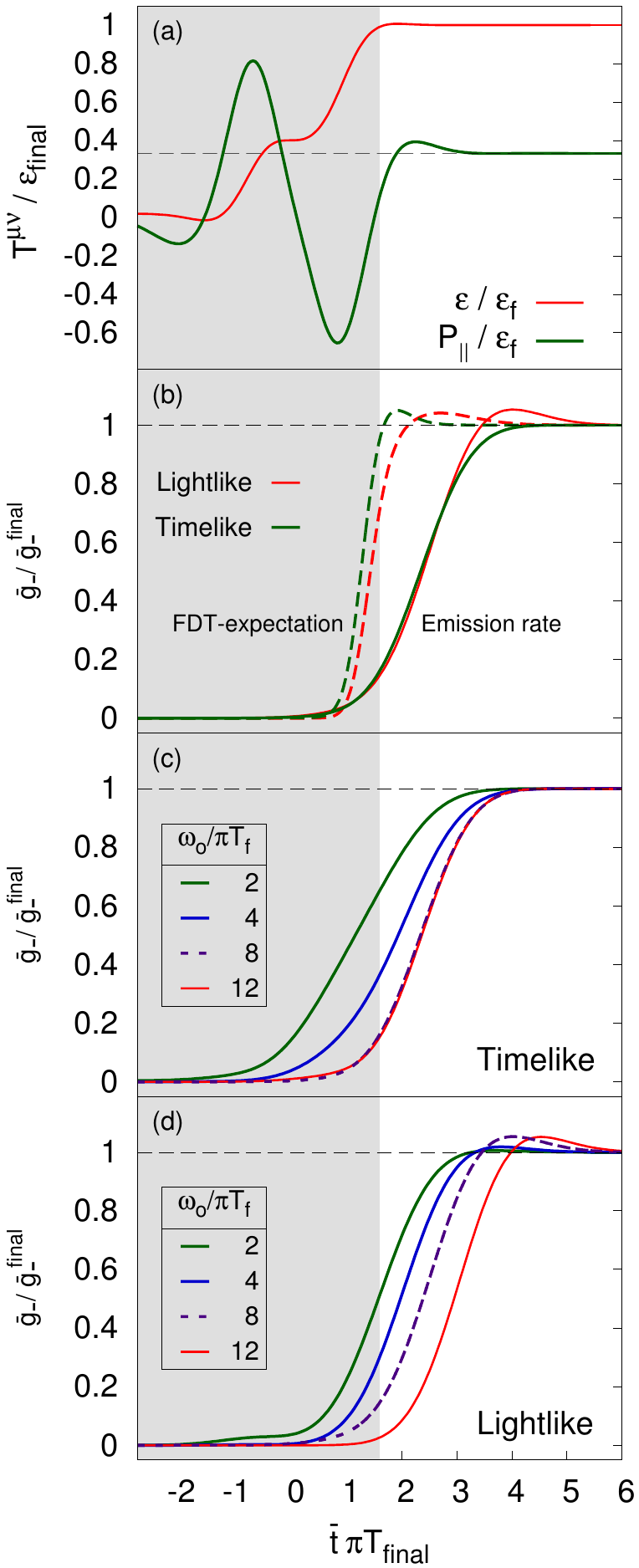}
\vskip -0.05in
\caption{(a) The SYM stress tensor $T^{\mu\nu}$ relative to the 
   final energy density ${\mathcal E}_{\rm f}$ as a function of time.
   The shaded band indicates
   when the energy density is changing  due to the work done 
   by the external gravitational field in the gauge theory. (b) 
   A non-equilibrium  emission rate, $\bar g_-(\bar t, q)/g^{\rm final}_-(q)$ (see \Eq{smeared_correlator}).  The emission rate is exhibited  for time-like momenta, with $\omega=8\pi T_{\rm f}$ and $\q=0$, and for light-like  momenta, with 
 $q_\perp=q_\parallel=\omega/\sqrt{2}$ and $\omega=8\pi T_{\rm f}$.
 The dashed lines show the FDT-expectation for the emission rate, {\it i.e.} the rate derived using the absorption rate $\bar g_+(\bar t, q)$ and the FDT result (\Eq{smearedFDT}).
 (c)  The emission rate for time-like momenta (as in (b)) 
 and various values of the
 frequency. (d) The emission rate for light-like momenta (as in (b)) at
 various values of the frequency.
\label{observables}
} 
\end{figure}

{\it{Discussion}}.---\Fig{observables}(a) shows a plot of the SYM stress 
$T^{\mu}_{\ \nu} = {\rm diag}( -\mathcal E, \mathcal P_{\perp}, \mathcal P_{\perp}, \mathcal P_{||})$
as a function of time. 
 In the distant past the stress is static and corresponds to a low temperature
equilibrium plasma of temperature $T_{\rm i}$.  When the 
boundary geometry (\ref{boundarygeometry}) starts to change, work is done on the system, the
energy density generally grows and the pressures oscillate.  
In \Fig{observables} the shaded region indicates that the energy density is still changing, and work 
is being done on the system by the external boundary source.  
After the energy density reaches its final value,  the transverse and longitudinal pressures $\mathcal P_{\perp}$ and  $\mathcal P_{||}$  equilibrate near time, $t \simeq  3/\pi T_{\rm f}$.

\Fig{observables}(b) shows the emission and absorption rates from
non-equilibrium plasmas for timelike and lightlike dilatons with $\omega
=8\pi T_{\rm f}$.  More precisely, the solid lines indicate Gabor transforms
of the emission rate $\bar g_-$ (eqs. (\ref{wigtrans}) and
(\ref{smeared_correlator})),  while the dashed lines indicate an estimate for
the emission rate, $\bar g_-^{\rm FDT}$, which uses the absorption rate $\bar
g_+$ and the coarse-grained FDT (\Eq{smearedFDT}).  From \Fig{observables}(b)
we see that that the emission rates equilibrate only after the stress tensor
and the absorption rates have  equilibrated.

The frequency dependence of the emission rates is studied in
\Fig{observables}(c) and (d).  The timelike emission rates in
\Fig{observables}(c) approach a universal curve at large frequency, indicating
that the emission of timelike dilatons equilibrates in a finite time as $\omega
\rightarrow\infty$.  By contrast, the lightlike emission rates in
\Fig{observables}(d) take an increasingly long time to equilibrate as $\omega
\rightarrow \infty$.


%

The qualitative features  of the equilibration process can be understood by
examining the  physics of non-equilibrium Hawking radiation recorded in
\Eq{bulk2bnd}. The Hawking flux from the black hole membrane can be considered
equilibrated when the horizon sources recorded by $\mathcal G^h_\pm$ satisfy
the horizon FDT \cite{deBoer:2008gu,Son:2009vu,CaronHuot:2011dr}.  
In general,  membrane equilibrium (as defined by the FDT) is reached after the background geometry and
boundary stress have stopped changing.  At this point, an
equilibrated Hawking flux is transmitted  to the boundary through the retarded
horizon-to-boundary Green functions in \Eq{bulk2bnd}.  Consequently, the
equilibration of the boundary two point functions is  causally delayed relative
to the boundary stress by the horizon-to-boundary transit time,  $\Delta
t\sim 2/\pi T$. This explains the  slight time shift between \Fig{observables}(a) and \Fig{observables}(b).

\begin{figure}
   \includegraphics[width=0.45\textwidth]{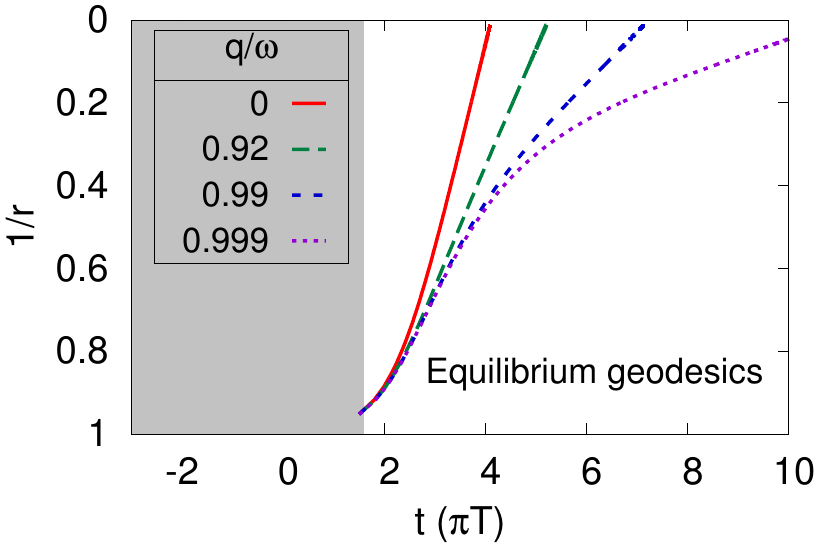} 
   \caption{ \label{geodesics}
      The geodesics in equilibrium which are relevant to the transport of Hawking
      radiation from the horizon to the boundary for a specified $|\q|/\omega$ \cite{Chesler:2008uy,Arnold:2011qi}.  (The horizon is at $r=\pi T = 1$ in the coordinates of \Eq{metric}.) Motivated
      by \Fig{observables}, the grey
      band indicates where  the non-equilibrium geometry  can not be reasonably modelled by equilibrium.
   }
\vspace{-0.2in}
\end{figure}

Similarly, the different equilibration patterns of  timelike and lightlike
modes  can be understood  with the dual physics of 
Hawking radiation.
The emission from the black hole membrane is the same for lightlike and timelike modes, since \Eq{horizoncorrelators} is independent of $\q$.  Thus, the timelike-spacelike
difference in the boundary theory is due to the horizon-to-boundary Green
functions. For large frequencies, the salient features of these Green
functions can be understood with geometric optics. In addition, for
coordinate time $\bar t \gsim 2.5{\pi T_{\rm f}}$ the metric is relatively
close to its equilibrium form, and equilibrium geodesics provide qualitative
insight to  the geometric optics of the full geometry.  \Fig{geodesics}  shows
a sample of $5d$ equilibrium geodesics that describe the transport of Hawking
radiation from the stretched horizon to the  boundary  for a specified
$|\q|/\omega$ in the boundary theory \cite{Chesler:2008uy,Arnold:2011qi}.

Examining \Fig{geodesics}, we see that the Hawking radiation which 
first arrives at the boundary is characterized by  high frequency, $\omega\rightarrow \infty$ 
with $\q$ fixed. These modes are transported
along null radial geodesics with a minimal transit time
of order $\Delta t\sim 2/\pi T$, and thus we expect timelike modes to thermalize
first.
Similar conclusions were also reached in
\cite{Balasubramanian:2010ce}. 
By contrast, the geodesics that determine the Green functions of hard  lightlike
modes  have $\omega \simeq |\q|$ large, and small virtuality $(w^2 - |\q|^2)/\omega^2 \sim 1/(\omega \sigma)$, where $\sigma=1/\pi T_{\rm f}$ is the smearing width in \Eq{smeared_correlator}. 
These geodesics (which are well known from computations of jet-energy loss   
in  strongly coupled plasmas \cite{Chesler:2008uy,Arnold:2011qi}) 
propagate up from the horizon and skim just below the boundary for a parametrically long time  compared to the final temperature,
$t  \sim (\omega \sigma)^{1/4}/\pi T_{\rm f}$. (This behavior is shown in \Fig{geodesics} for the equilibrium geometry.) 
Thus, lightlike boundary
correlators at late times,  $t \lsim  (\omega\sigma)^{1/4} /\pi T_{\rm f}$,
reflect the emission of non-equilibrium  Hawking radiation  at earlier times $t \lsim 1/\pi T_f$. The delayed equilibration seen in \Fig{observables}(d) is 
consistent with a $(\omega\sigma)^{1/4}$ expectation.


Finally, we note two marked differences between the evolution of
quantum fluctuations at weak and strong coupling.  First, note that
the back-reaction between the fluctuations and the metric was neglected in
this study.
At weak coupling this approximation is of limited
use, since quantum fluctuations seed instabilities which grow exponentially,  and
invalidate the approximation scheme before the plasma thermalizes \cite{Mrowczynski:1993qm, Arnold:2003rq, Rebhan:2004ur, Romatschke:2005pm,Dusling201169}.  In the
gravitational theory, however, the timescale for growth is the same as the
equilibration timescale of the background metric.  Thus, in the SYM plasma a
nascent instability can grow by at most one $e-$folding before being damped by
the equilibrium black hole metric.  
Second, note that the fluctuations equilibrate even in the
collisionless (or linearized) approximation  that we are using.  In weakly
coupled (but strong field) plasmas the analogous Hartree approximations fail
satisfy the FDT at late times \cite{Berges:2004yj,Dusling201169}.  These  marked differences with weakly
coupled  plasmas provide  a useful foil to perturbative
analyses of the thermalization in heavy-ion collisions. 

In summary, we have computed the Hawking flux from non-equilibrium black holes.
From a gravitational perspective, the computation is both conceptually and
technically challenging. The physics of this
novel emission process determines the qualitative features of the
thermalization of fluctuations in strongly  coupled plasmas.

{\it{Acknowledgments}}.---PMC is supported by a Pappalardo Fellowship 
in physics at MIT.  DT is supported in part by the Sloan Foundation and by the 
Department of Energy through the Outstanding Junior Investigator program, DE-FG-02-08ER4154.

\bibliography{refs}%
\end{document}